\documentclass[a4paper]{jpconf}
\usepackage{graphicx}
\usepackage{color}
\usepackage{latexsym}
\usepackage{amsfonts}
\usepackage{multirow}

\newcommand{\be}{\begin{eqnarray}}
\newcommand{\ee}{\end{eqnarray}}

\begin{document}

\title{Testing general relativity with X-ray reflection spectroscopy of MCG-06-30-15}

\author{Ashutosh Tripathi}

\address{Center for Field Theory and Particle Physics and Department of Physics, Fudan University, 200438 Shanghai, China}

\ead{ashutosh\_tripathi@fudan.edu}

\begin{abstract}
The spacetime geometry around astrophysical black holes is well approximated by the Kerr metric, but deviations from standard predictions are possible in a number of scenarios beyond Einstein's gravity and in the presence of exotic matter. In this paper, we present the constraints on possible deviations from the Kerr solution using X-ray reflection spectroscopy from the analysis of real data. We use the relativistic X-ray reflection code RELXILL modified to a generic stationary, axisymmetric and asymptotically flat black hole metric. We analyze 350 ks long XMM-Newton observations of the AGN in MCG-06-30-15 taken during July-August 2001 and constrain the Johannsen deformation parameter $\alpha_{13}$.
\end{abstract}


\section{Introduction}
The general theory of relativity was proposed by Albert Einstein in 1915 and it has been used to describe the dynamics of spacetime since then. Several experiments, like binary pulsar timing array, calculation of light deviation by Sun etc., has been performed in order to test the general relativity in weak gravity regime~\cite{will}. The theory is standard in the sense that the theoretical predictions from the theory agree well with experimental deductions. In order to test deviation from general relativity in describing the dynamics of spacetime, the theory needs to be tested in strong gravity regime. In other words, one need to design an experiment to test this theory where gravity is very strong.
        The ideal probe in the universe to test general relativity in strong gravity regime are black holes~\cite{book}. \\
        
        In Einstein gravity, an uncharged black hole is described by Kerr solution and, only two parameters, namely, mass (M) and angular momentum (J), are required for its complete quantification~\cite{hair}. Although Kerr metric should describe a black hole, deviations from it are also possible. It is possible that deviations are radiated away during initial stage of black hole formation in the form of gravitational waves . 
        
There are two distinct methods through which general relativity can be tested; Gravitational and electromagnetic radiation~\cite{review, review2, review-gw}. Electromagnetic approach asssumes the geodesic motion of Kerr metric for analysis and can be easily generalized to test the deviation~\cite{review, review2}. This approach analyze the properties of electromagnetic radiation from the accretion disk close to black hole. It depends on the motion of gas in strong gravity region and photon propagation from the emission point in the disk to the faraway region where spacetime is almost flat.
        
        There are two leading electromagnetic techniques for testing the Kerr metric hypothesis; continuum fitting method and the reflection method~\cite{book}. Reflection method involves the analysis of relativistically smeared reflection spectrum of thin accretion disk. It can be applied to both stellar mass black holes and supermassive black holes. It is independent of black hole mass and distance. The inclination angle of the disk w.r.t. to line of sight is calculated by fitting the reflection spectrum. For continuum method, these three quantities must be inferred from other methods.
        
         In this work, we employ a test metric, Johannsen metric~\cite{Johannsen:2015pca}, which is a generalized form of Kerr metric to describe spacetime around black hole. The deviations from Kerr metric are quantified by deformation parameters which are constrained using astrophysical data. Here, we use XMM-Newton 2001 observations of the supermassive black hole MCG-06-30-15.


\section{The source MCG-06-30-15}
MCG-06-30-15 is a narrow line seyfert 1 (NLSy1) galaxy situated at a distance of 37 Mpc ($z =$ 0.008). 
This source has shown significant deviation 
from simple power-law based on the evidence from both absorption and emission feature from cold iron line~\cite{s-1}. 
The first X-ray observation of this source was reported with \textit{ASCA}, which was taken during 1993 July-August~\cite{s-2}. 
The warm absorber edges and iron emission lines are clearly resolved, for the first time, using this observation. 
They, also, discovered large change in column density between two observations taken during this period. 
The detection of relativistic effects in X-ray emission line ($K_\alpha$ line) from ionized iron 
in this source was first reported in~\cite{s-3}. 
The observed K$_\alpha$ line is broad and antisymmetric because of the relativistic effects, 
which are so strong that it was believed to be originated from the innermost region of accretion disk.
Iwasawa et al. ~\cite{s-4} analysed the variability of the source during \textit{ASCA} observation. The peculiar 
line profile during 'deep minimum state' suggests that the line emitting region is very close to a central 
spinning black hole where enormous gravitational effect operates. 

 Multi-wavelength study of this source was explained in ~\cite{s-5}, which confirms the presence of double zone warm absorber. 
The analysis at softer energies is presented in ~\cite{s-6} for an energy range of 0.32-1.7 KeV and they found a reasonable fit using multizone dusty 
warm absorber. 

Brenneman \& Reynolds presented the analysis of iron line and warm absorber of 2001 XMM-Newton observation in ~\cite{i2}. Their new model, {\sc kerrdisk},   allows black hole spin to be a free parameter and found the value of spin to be $0.989\pm0.002$. 
The spectral variability was examined in ~\cite{s-7} using simultaneous XMM-Newton and NuSTAR observations taken in 2013. 
The extreme variability of the source can be explained in terms of intrinsic X-ray fluctuations and light bending model.
 
 The relativistic broadening of iron line in this source make it suitable for studying the deviation from Kerr metric.


 \section{Observations and Data Reduction}
 
 MCG-06-30-15 was observed by XMM-Newton ~\cite{s-8} on 31 July, 2 August and 4 August 2001 for a total period of 350 ks during 
 three revolutions 301,302 and 303 respectively. The observations were taken with EPIC pn and MOS1/2 cameras operated in 
 small window and medium filter mode~\cite{s-9}. Data from MOS detectors are not analyzed because they suffered significantly 
 from pile-up. 
 
 We used {\it XMM-Newton} Science Analysis Software (SAS) version 16.0.0 for data reduction.
 Three EPIC pn fits files of different observation IDs were combined to form one single fits file using ftool {\it MERGE}.
The good time intervals (GTIs) were generated using \textit{TABTIGEN} which contains data free of proton flares.  
The processed event files were generated by combining GTI and event list files and, then, filtered using 
single and double events (PATTERN $\leq$ 4) but excluding events that are at the edge of 
CCD (FLAG=0). To obtain source event files, we have extracted an area of radius 40 arcsec. 
For background events, an area of radius 50 arcsec, located as far away as possible from the source, 
is chosen to avoid any of its contribution. After backscaling source and 
background spectra, response files and ancillary files were generated using rmfgen and arfgen, respectively.
Lastly, In order to oversample instrumental resolution by a factor of 3 and to have minimum counts of 30 
per background-subtracted spectral channel so that $\chi^2$ statistics can be used, the spectra were rebinned using {\it SPECGROUP}.


 \section{Theoretical Models}

\begin{figure}[t]
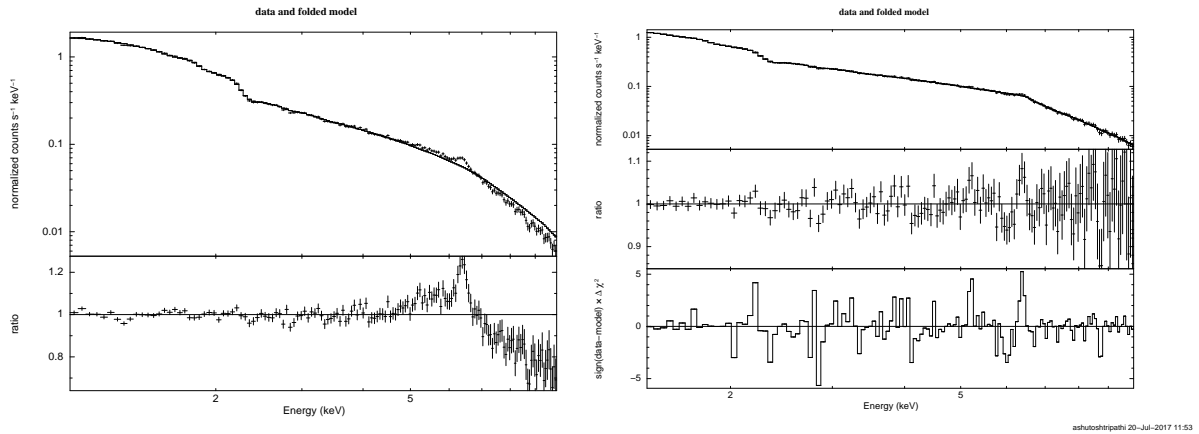

\includegraphics[scale=0.3,angle=270]{tripathi_fig1a.ps}
\includegraphics[scale=0.3,angle=270]{tripathi_fig1b.ps}
\caption{{\it Left panel} : The 2001 XMM-Newton observation ratioed against the power law. {\it Right Panel} : The data to model ratio for the model employed in this paper. }
\end{figure}

 In figure 1, we show the ratio of data to power law continuum in order to identify different features in the data, which we have tried to analyze with our theoretical models in the present work. The source is highly variable throughout the entire energy range. Most prominent feature is the emission line at 6.4 KeV, which is likely due to iron $K_\alpha$ emission. \\
 In order to model the variability in soft energy range ($<2.0$ KeV), we employ double zoned warm absorber. Physically, warm absorbers are ionized absorbing gas believed to be originated from accretion disk. It surrounds the central engine of AGN like a torus (see for e.g.~\cite{i2} and references therein). They contain a rich forest of lines and edges from various species of gas and dust. They incorporate several "zones" of materials, distinct in their kinematic properties as well as their column densities and ionizations, but maintained in pressure balance. For computation, it is approximated as discrete zones characterized by a column density and ionization parameter. {\sc xstar} spectral synthesis package for photoionized gases is used to construct a grid of warm absorbers as a function of column density and ionization parameter. They are multiplicative models, absorber models that can be applied to any emission spectrum.\\
 
 The disk produces a whole spectrum of recombination and flourescent lines. To obtain realistic constraints, we need to model the entire X-ray reflection spectrum (see, e.g., \cite{book} and references therein). For X-ray reflection, {\sc xillver} provides the best treatment of radiative transfer using photoionization routines from {\sc xstar} which contains the most detailed atomic database for modelling photoionized X-ray spectra. {\sc xillver}  provides a detailed treatment of K-shell atomic properties of ionized ions. The most prominent one is iron $K_\alpha$ line at 6.4 KeV. This line is narrower in emitter's reference but it becomes broad and skewed in receiver's frame of reference because of general relativistic effects (Doppler effect, gravitational redshift) which are prominent in strong gravity regime. These effects are so strong that the emission lines can possibly be radiated from innermost regions of black hole. \\

 {\sc relconv} is a relativistic convolution code assuming Kerr metric. It gives the spectrum measured by a distant observer given the local spectrum at any emission point in the disk. The code {\sc relxill} is the combination of {\sc relconv} and {\sc xillver}. It describes relativistic smearing of X-ray reflection spectrum near the black hole~\cite{rel1,rel2,rel3}. {\sc relxill\_nk} is the extension of {\sc relxill} to Johannsen metric, of which Kerr metric is a special case~\cite{relnk}.  In order to test Kerr metric, {\sc relxill\_nk} must describe the X-ray reflection spectrum in a background metric, more general than Kerr metric, which contains some deformation parameters in addition to mass (M) and angular momentum (J). 
 
\begin{figure}[t]
\centering
\includegraphics[scale=0.6]{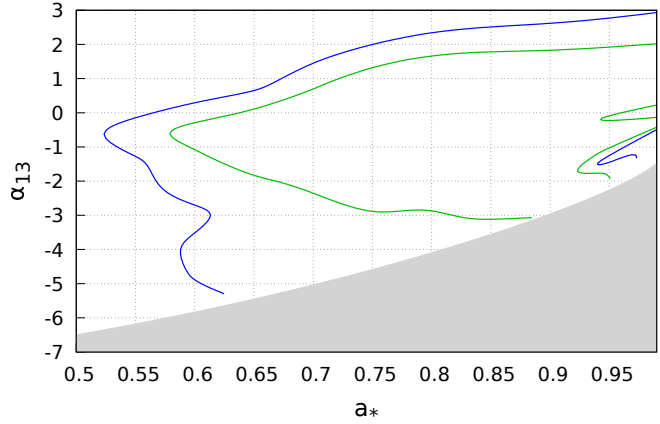}
\caption{90\% and 95 \% confidence contours in $a^* - \alpha_{13}$ plane. }
\end{figure}

\begin{table}
\begin{tabular}{cclc}\hline 
Components&Parameters&Description&Best fit value\\ \hline
TBABS& $N_H$ ($10^{22}$) & Column density for Galactic absorption&0.29 $\pm$0.06\\ \hline
\multirow{2}{*}{WARMABS1}&$N_H$ ($10^{22}$) & Column density for I warm absorber &0.24 $\pm$0.11\\
&$log\xi$ &Ionization parameter for I warm absorber &1.40 $\pm$0.34\\ \hline
\multirow{2}{*}{WARMABS2}&$N_H$ ($10^{22}$) & Column density for II warm absorber&0.40$\pm$0.27\\
&$log\xi$&Ionization parameter for II warm absorber & 2.47 $\pm$0.27\\ \hline
\multirow{8}{*}{RELXILL\_NK}&$Index1$ &Emissivity index for $r<Rbr$& 2.3$\pm$0.28\\
&$Index2$& Emissivity index for $r>Rbr$& $3.0^*$\\
&$R_{br}$& Break radius& 76.60 $\pm$ 121.18\\
&$a$&spin of Black hole&0.92$\pm$0.12 \\
&$i (deg)$ & Inclination angle w.r.t. line of sight& 3.6 $\pm$35.8\\
&$\Gamma$   & Power law Index of the primary source ($E^{-\Gamma}$)& 1.98$\pm$0.03\\
&$log(\xi_{refl})$   & Ionization parameter at the inner edge of disk& 3.34$\pm$0.04\\
&$Afe$  & Iron abundance in solar units&$2.21^*$\\
&$refl\_frac$ & Reflection fraction&1.72$\pm$0.57\\
&$defpar\_value$ & Deformation parameter $\alpha_{13}$&-0.82$\pm$0.08\\ \hline
\end{tabular}
\caption{Best fit parameters of different components of the model used and their description. The parameters marked with * are frozen during analysis.}
\end{table}

For modelling spectrum and variability, following model is used to fit the observation in {\sc XSPEC}
 \be\label{eq-m3}
&& {\rm Model~1 : \,\,\, {\sc TBABS*WARMABS1*WARMABS2*RELXILL\_NK}} \, . 
\ee
{\sc relxill\_nk} models power law continuum and reflection component from region near black hole and take into account the
relativistic smearing of iron line. The cut-off energy of input spectrum is taken to be 300 KeV. The inner edge of the disk is assumed to be located at the radius of innermost stable circular orbit ($R_{isco}$). {\sc warmabs1} and {\sc warmabs2} are two warm absorbers used to model ionized gas around the source. {\sc tbabs} models the galactic absorption.


\section{Results and Discussion}

 In table 1, we report the best fit parameters for the model which is used to fit the spectrum in the energy range of 1.3-10.0 KeV. The reduced chi square value ($\chi^2_{red}$) is found to be 1.04. The value of spin is found to be 0.92 which is consistent with earlier claims of this source being a rapidly spinning black hole. Due to degeneracies of iron-line model parameters, some parameters could not be constrained. Figure 2 shows $90\%$ and $95\%$ confidence contours in spin- $\alpha_{13}$ plane. The Kerr metric is recovered for this model as the null value of deformation parameter is included in confidence ellipses. However, it could not constrain the spin effectively. This is because of high variability of the source and the presence of complex absorption and emission features, which could not be modelled properly with warm absorber and galactic absorption.  \\


\section{Conclusions and Future Work}

We tested the deviations from general relativity using astrophysical data. This is achieved by modifying the existing X-ray reflection model, which assumes Kerr metric, to a more generic metric. We obtained constraints on deformation parameters by analyzing 350 ks long XMM-Newton observation of MCG-06-30-15 which is famous for its relativistically broadened iron line. Unfortunately, the current model cannot constrain the parameters properly because of highly variable nature of the source. Other sources can also be analyzed in order to have better constraints on deformation parameters~\cite{zheng}. Future X-ray missions like ATHENA and LAD/eXTP~\cite{snzhang} can be proved useful in constraining possible deviations from Kerr metric.


\ack

This research is based on observations obtained with {\it XMM-Newton}, an ESA science mission with instruments and 
contributions directly funded by ESA Member States and NASA.

AT acknowledges support from the China Scholarship Council (CSC), grant no.\  2016GXZR89.


\end{document}